%% file: cancer.tex
\documentclass[format=acmsmall, review=false, screen=true]{acmart}

\usepackage{booktabs} 

\usepackage[ruled]{algorithm2e} 

\SetAlFnt{\small}
\SetAlCapFnt{\small}
\SetAlCapNameFnt{\small}
\SetAlCapHSkip{0pt}
\IncMargin{-\parindent}

\setcopyright{acmlicensed}
\acmJournal{PACMHCI}
\acmYear{2018} \acmVolume{2} \acmNumber{CSCW} \acmArticle{58} \acmMonth{11} \acmPrice{15.00}\acmDOI{10.1145/3274327} 

\received{April 2018}
\received[revised]{July 2018}
\received[accepted]{September 2018}

\newcommand\Tf{\rule{0pt}{2.8ex}} 
\newcommand\Ts{\rule{0pt}{2.2ex}}
\newcommand{\n}[1]{\textcolor{black}{#1}}

\begin{document}
\title[Fake Cures: User-centric Modeling of Health Misinformation in Social Media]{Fake Cures: User-centric Modeling of Health Misinformation in Social Media}

\author{Amira Ghenai}
\orcid{0000-0001-5583-3008}
\affiliation{%
  \institution{University of Waterloo}
  \streetaddress{200 University Ave W}
  \city{Waterloo}
  \state{ON}
  \postcode{N2L 3G1}
  \country{Canada}}
\email{aghenai@uwaterloo.ca}

\author{Yelena Mejova}
\orcid{0000-0001-5560-4109}
\affiliation{%
  \institution{ISI Foundation}
  \streetaddress{Via Chisola, 5}
  \city{Torino}
  \state{TO}
  \postcode{10126}
  \country{Italy}}
\email{yelenamejova@acm.org}

\begin{abstract}
Social media's unfettered access has made it an important venue for health discussion and a resource for patients and their loved ones. However, the quality of the information available, as well as the motivations of its posters, has been questioned. This work examines the individuals on social media that are posting questionable health-related information, and in particular promoting cancer treatments which have been shown to be ineffective (making it a kind of misinformation, willful or not). Using a multi-stage user selection process, we study 4,212 Twitter users who have posted about one of 139 such ``treatments'', and compare them to a baseline of users generally interested in cancer. Considering features capturing user attributes, writing style, and sentiment, we build a classifier which is able to identify users prone to propagate such misinformation at an accuracy of over 90\%, providing a potential tool for public health officials to identify such individuals for preventive intervention.
\end{abstract}

%
%



\maketitle

\renewcommand{\shortauthors}{A. Ghenai \& Y. Mejova}

\input{introduction}

\input{relatedwork}

\input{data}

\input{results}

\input{discussion}

\input{conclusion}

\section{Acknowledgements}

We would like to express our gratitude to Dr. Jeremie Arash Rafii Tabrizi who has assisted in misinformation topic evaluation, query validation, and overall advise.

\bibliographystyle{ACM-Reference-Format}
\bibliography{cancer}
\end{document}

%% file: introduction.tex
\section{Introduction}

Searching and sharing health information online is becoming a common practice. A 2011 survey indicated that as many as 62\% of adult Internet users in U.S. used social network sites for health related topics, from following friends' health experiences or updates to gathering health related information \cite{fox2011social}. Some manage their health via general platforms such as PatientsLikeMe, while others join condition-specific communities like TuDiabetes, yet others share their experiences in general-purpose social media \cite{de2014seeking,korda2013harnessing}. For instance, Paul et al. \cite{paul2011twitter} showed that a significant number of personal and health-related questions were being asked on the microblogging platform Twitter, which is becoming the top destination for both patients and healthcare professionals \cite{antheunis2013patients}. The social component of such interactions is especially important in educating the population on health matters, as individual opinions can be strongly biased by their social network \cite{lau2011online}.


While social media use for health management is growing, so do the concerns over the lack of accountability, dubious quality and loose confidentiality \cite{greene2011online,moorhead2013new}. With few legal constraints imposed on often profit-seeking websites, social media provides a dynamic forum for propagating possible medical misinformation \cite{frish2017social}. Recent rise in vaccine hesitancy has been linked to an active movement on Twitter, promoting conspiratorial thinking and mistrust in the government \cite{mitra2016understanding}. Image sharing platforms such as Flickr and Instagram have become battlegrounds between the pro-anorexia movement and physicians attempting to intervene \cite{yom2012pro,chancellor2016thyghgapp}. Uncertainty surrounding infectious disease outbreaks, such as the Zika epidemic of 2016, yielded rumors and speculations about its causes, preventive measures, and consequences \cite{dredze2016zika,Ghenai2017catching}. 

In this study we turn to the individuals sharing questionable medical information on Twitter, in particular cancer treatments which have been medically proven to be ineffective. \n{Having around 336 million monthly active users in the first quarter of 2018\footnote{\url{https://www.statista.com/statistics/282087/number-of-monthly-active-twitter-users/}}, Twitter is one of the largest social media websites expressly dedicated to the sharing of information, including that on cancer.} Compiling hundreds of thousands of tweets on 139 queries spanning acupuncture, cinnamon, reflexology, and vitamin C, we apply strict selective criteria employing human/organization classification \cite{McCorriston2015Organizations}, name dictionaries, usage thresholds, and crowdsourced \n{relevance} refinement resulting in 4,212 users, which we then compare to those mentioning cancer in general from a previous study \cite{paul2014discovering}. Employing previous research on rumor detection, we characterize these users in multi-faceted feature spaces, encompassing user attributes, linguistic style, sentiment, and post timing. We find users who have a more sophisticated language, who are interested in cancer, but who are not personally involved with the illness. We build a logistic regression model which, out of Twitter users mentioning cancer, is able to identify those who will eventually post a piece of misinformation with a high level of accuracy.

Misinformation on social media is an urgent issue, and even more so in the health field. \n{This paper is one of the first to look into the characteristics of users propagating unverified ``cures'' of cancer on Twitter as a case study of tracking health misinformation outside crisis communication management domain. The identification of potential sources of such misinformation would allow public health officials to monitor social media discourse, characterize the deficiencies in current communication strategy, and detect new misinformation before it causes serious harm.} 

%% file: relatedwork.tex
\section{Related Work}
\label{related work}

The present work marries two burgeoning directions of social media research: the tracking of misinformation and the measurement of health-related attitudes and activities. Below we describe the latest developments in both topics, as well as the recent attempts at tracking health misinformation in particular.

\textbf{Misinformation tracking.} In 2017, the term ``fake news'' has been named Collins' Word of the Year, referring to ``false, often sensational, information''\footnote{\url{https://ind.pn/2AnI2Bw}}. In the context of news, rumors and misinformation have been associated with political sphere, with the latest works proposing data mining solutions \cite{shu2017fake} which encompass opinionated language detection \cite{chen2015misleading}, visual feature extraction \cite{gupta2013faking} and user group characteristics \cite{ma2015detect}. In fact, users play an important role, as misinformation sometimes originates from automated accounts working in synchrony -- bot nets -- with \cite{shao2017spread} claiming millions of political tweets spread thusly during and following the 2016 U.S. Presidential election. Tools to detect such bots include Hoaxy \cite{shao2016hoaxy} to track the spread of claims and Botometer \cite{davis2016botornot} to detect social bots. Beyond the automated accounts, case studies of incidents such as the Ukrainian conflict \cite{khaldarova2016fake} and the mass shootings in US \cite{starbird2017examining} examine human reactions to questionable information online. The provenance and motivation behind such information has increasingly become a contentious issue, as speculations rose that important political decisions, including the United Kingdom vote to leave the European Union and the election of Donald Trump to the U.S. Presidency, have been potentially swayed by forces outside those nations \cite{hern2017howsocialmedia}, posing a danger to democracy itself.

\textbf{Health-related attitudes on social media.} Beyond politics, social media also provides ample resources for health-related decision making,  capturing behaviors and attitudes impacting individual health. Automated methods have been devised for tracking marijuana use on Twitter \cite{ginart2016drugs}, and to capture attitudes toward legal drugs including Xanax and Adderall \cite{seaman2016prevalence}. Forum threads have been analyzed by Wu et al. \cite{yang2013harnessing} to discover adverse drug effects and drug interactions, using association mining. Likewise, behaviors related to lifestyle diseases such as diabetes type 2 and obesity have been tracked using Twitter (a microblogging platform) \cite{abbar2015you}, Instagram (a photo sharing platform) \cite{mejova2015dietary}, and Facebook (a social network) \cite{araujo2017using}, along with attitudes toward food and diet \cite{mejova2016fetishizing}. A study of a community promoting anorexia on Flickr (another photo sharing platform) \cite{yom2012pro} showed that attempts of the anti-anorexia programs to infiltrate the community with intervention messages tagged with pro-anorexia tags was counterproductive in the long run (with users exposed to such remaining in the group longer). Further uses of social media to gauge the efficacy of health communication includes a recent study of breast cancer mammography advisory on Twitter \cite{nastasi2017breast} which found many users to be confused by it than to approve of it. Finally, public awareness of health-related topics has been recently gauged through the advertising platforms provided by these social media -- for instance using Facebook Advertising Manager to estimate the number of Facebook users interested in diabetes-related topics \cite{araujo2017visualizing}. Thus, as social media adoptions increases, so does the health-related discussion and information seeking on these platforms, allowing for large-scale analysis and tracking.

\textbf{Health misinformation online.} For over a decade, medical sociologists have studied Internet as a new component of health ecosystems, with especially cancer patients utilizing it to collect information and make treatment decisions \cite{chen2001impact}. The patient reliance on the Web has resulted in a patient--Web--physician ``triangulation'' \cite{wald2007untangling} with benefits such as more efficient use of clinical time and additional support from online support groups, coupled with potential harms like the dangers posed by the variable quality of information, unnecessary visits to a physician, and exacerbating existing socioeconomic health disparities. Early on, the use of Internet by patients has been shown to be problematic \cite{schmidt2004assessing}. For instance,  \cite{dy2012effect} find that the quality of Web search results varies for differently worded medical queries. Democratization of content publishing may also be exacerbating quality concerns, as YouTube videos have been found to contain instances of public display of harmful or unhealthy behaviors, promotion of tobacco to consumers, and distorting policy and research funding agendas \cite{lau2012social}. Moreover, concentrated efforts promoting doubt of medical establishment, such as the ``anti-vaxxer'' movement on Twitter, play into the larger skepticism of government and conspiratorial thinking \cite{mitra2016understanding}.

Recent attempts to track health misinformation include ``VAC Medi+ board'', an online interactive visualization framework integrating heterogeneous real-time data streams with Twitter data \cite{kostkova2016vac}, which tracks the spread of vaccine related information on Twitter and the sources of information spread. Such social media content promoting vaccine hesitancy has been shown to impact future opinions of its users, and potentially their subsequent behaviors \cite{dunn2015associations}. It is also possible to study the impact of the interventions designed to change such behaviors, such as in the previously mentioned study by \cite{yom2012pro} on the potentially unsuccessful attempts to infuse anti-anorexia content into pro-anorexia communities on Flickr. Dynamic nature of social media also allows for fast spread of misinformation during an ongoing epidemic, and machine learning techniques have been deployed to track such content, for instance during the 2016 Zika outbreak \cite{dredze2016zika,Ghenai2017catching}. To help social media users, tools are being developed to ease the verification of health claims via natural language processing and retrieval of necessary medical literature \cite{samuel2018medfact}.

However, little attention has been paid to the modeling and understanding of the spread of cancer ``complementary and alternative medicine'' (CAM) on the Internet. A 2008 survey of 80 cancer patients found that, when going online, respondents dealt with emotional stress of being reminded about their prognosis, and were seeking second opinion of a doctor before using CAM promoted online \cite{broom2008role}. However, more recent surveys find internet to be increasingly important source of information on CAM, with around half of patients using the alternative medicines, as well as their relatives, getting their health advise online \cite{huebner2014user,ebel2015perception}. Thus, in this work we focus on the kinds of individuals who are susceptible to propagating unverified information about cancer treatments which have been found to be ineffective at treating cancer. Guided by previous literature on rumors (see \cite{rosnow2005rumor} for an overview), we examine writing style, personal involvement, and other interests, as described below, in order to model individuals particularly vulnerable to such misinformation.




%% file: data.tex
\section{Data Collection}

Dataset used in this work consists of tweets belonging to two groups of users: (1) a ``rumor'' group who have posted content promoting one of 139 cancer ``treatments'' which have been proven ineffective, and (2) a ``control'' group who posted generally about cancer, but not on any of the above topics. The initial data gathering, and multiple steps of user selection and relevance refinement are described below.

\subsection{Health Rumor and Control Data Collection}

As the focus of this study is the behavior of users who post on social media health content of questionable nature, we begin by compiling a set of purported cancer ``cures'' which have been shown by experimentation and medical professionals to be ineffective. Four of such dubious cancer treatments come from White \& Hassan \cite{white2014content} where authors judged and reached a consensus about the medical treatments' efficacy by reading the corresponding Cochrane Review \cite{cipriani2011cochrane,higgins08:cochrane}. Next, we collect nine rumor topics from David Colquhoun (Professor of Pharmacology at University College London) blog\footnote{\url{http://www.dcscience.net/}}. Professor David's blog focuses particularly on alternative medicine such as homoeopathy, traditional Chinese medicine and herbal medicine. 
Finally, we collect 126 unproven cancer treatments listed in the ``List of unproven and dis-proven cancer treatments''\footnote{\url{https://en.wikipedia.org/wiki/List\_of\_unproven\_and\_disproven\_cancer\_treatments\#Ineffective\_treatments}} Wikipedia page which was refereed by Cancer Research UK\footnote{\url{http://scienceblog.cancerresearchuk.org/2014/03/24/dont-believe-the-hype-10-persistent-cancer-myths-debunked/\#superfoods}}. The selection of these unproven cancer treatments is then supervised by a trained oncologist (acknowledged below) in order to validate the ground truth of the treatments' efficacy, making sure all collected ``treatments'' are indeed ineffective. This process results in a total of 139 cancer treatment-related topics\footnote{The topics, along with the keyword queries are available at \url{https://tinyurl.com/y78mkg6s}}. Henceforth we will call these topics cancer treatment rumors, or simply rumor topics. Note that some of the above treatments may be effective in alleviating some of the symptoms of cancer, but do not actually affect the underlying progression of cancer (see Discussion for more).

\begin{figure}
    \centering
    \includegraphics[width=0.7\linewidth]{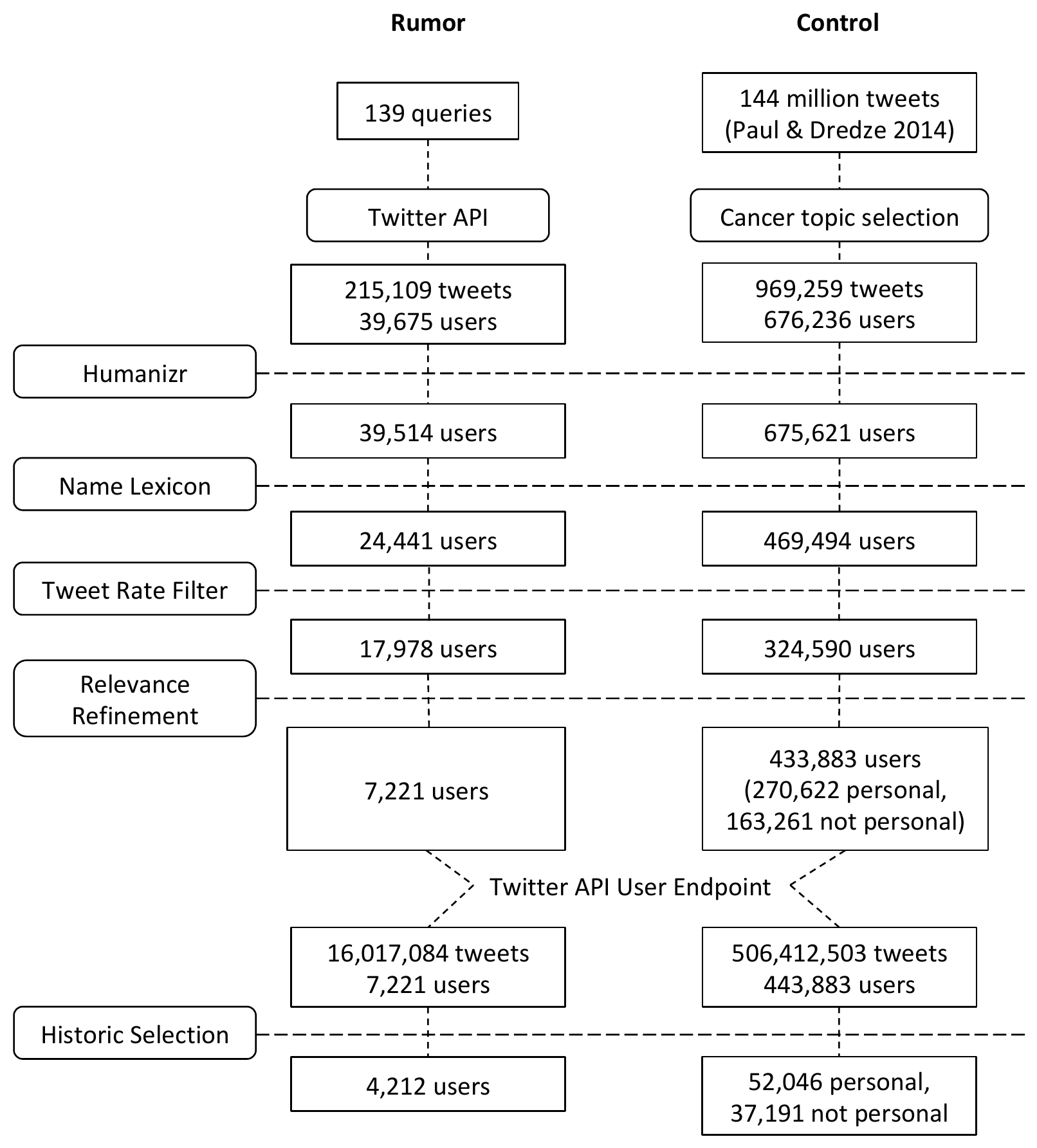}
    \caption{Data collection and refinement process.}
    \label{figuredatarefinement}
\end{figure}

Considering Twitter users posting about the above topics as the ``rumor'' group, we turn to existing research on health discussions for the ``control'' group. These would be people talking in general about cancer such as cancer causes, prevention, symptoms, and awareness or sharing personal experiences with the medical condition. For this purpose, we use Paul \& Dredze~\cite{paul2014discovering} public health topics dataset which consists of 144 million tweets that are related to a selection of health topics gathered during the period of 01 August 2011 - 28 February 2013. As the focus of this study is cancer, we focus on the 676,236 users who have posted 969,259 tweets in this dataset (for a summary of user selection process, see Figure \ref{figuredatarefinement}).

Next, we turn back to the Rumor group and collect tweets on rumor topics that span the same time period as the control. For every rumor topic, we hand craft a query and expand it using general domain tools such as Google search and Google keyword planner\footnote{\url{https://adwords.google.com/ko/KeywordPlanner}} as well as medical domain tools including Mayo clinic\footnote{\url{http://www.mayoclinic.org/}}, Merriam-Webster dictionary\footnote{\url{https://www.merriam-webster.com/}} and SNOMED CT BioPortal which is a repository of biomedical ontologies \cite{whetzel2011bioportal}. For instance, below is an expanded query for topic \emph{shark cartilage}, which has been shown to have no effect on survival rate or quality of life for cancer patients \cite{loprinzi2005evaluation}:

\begin{quote}
{\scriptsize \texttt{``Shark cartilage'' OR ``AE-941'' OR ``Marine Collagen'' OR ``Marine Liquid Cartilage'' OR ``MSI-1256F'' OR ``Neovastat'' OR ``Sphyrna lewini'' OR ``Squalus:acanthias'') AND cancer}}
\end{quote}

It includes a typical way to refer to the topic, as well as more technical version of the treatment, and related products such as Neovastat, a shark cartilage extract\footnote{\url{https://www.cancer.gov/publications/dictionaries/cancer-drug?cdrid=42021}}. Once again, the extended queries were verified by an oncologist for correctness and completeness. Using the Twitter Streaming Application Program Interface (API), we collect a total of 215,109 tweets about these rumor topics (see Figure \ref{figuredatarefinement}) spanning \n{2011-2013} and 39,675 users. 

\subsection{User Selection}

For both rumor and control tweets, we aim at eliminating users that are not human such as bots, organizations, or whose tweets do not refer to the actual topics of interest (but were picked up due to a faulty or ambiguous keyword matching). We perform several steps to raise the likelihood the selected users meet the above criteria. 

\begin{itemize}

\item We apply the \textit{Humanizr} tool \cite{McCorriston2015Organizations} to the tweets, which was shown to have an accuracy of 94.1\% predicting whether a Twitter user is an organization account. In this step, we remove 161 accounts from the rumor user set and 615 from the control set. 

\item Next, we compile a (human) name dictionary with associated genders by combining names extracted from a large collection of Google$+$ accounts \cite{magno2014international} with baby names published by National Records of Scotland\footnote{\url{https://www.nrscotland.gov.uk/statistics-and-data/statistics/statistics-by-theme/vital-events/names/babies-first-names/full-lists-of-babies-first-names-2010-to-2014}} and United States National Security\footnote{\url{https://www.ssa.gov/oact/babynames/limits.html}}, resulting in a dictionary containing 106,683 names. After matching this dictionary to user names, as well as applying heuristics (such as having ``Mrs." or ``Mr."), we keep only users with \n{a matching name or identifier}, excluding 15,164 (38.2\%) users from rumor and 207,394 (30.6\%) from control sets. As illustrated in Table \ref{table:gender}, \n{name matched} accounts are more often verified accounts, have fewer overall tweets, followers and following users than the non-gendered users, indicating they are less active than \n{those not matching a name in our dictionary}.

\begin{table}[t]
\centering
\caption{Average statistics of users whose names were found in \n{name dictionary versus those not found}.}\vspace{2mm}
\label{table:gender}
\begin{tabular}{lrrrr}
\hline
\textbf{Name match?}\Ts & \textbf{Followers} & \textbf{Followees} & \textbf{Tweets} & \textbf{Verified} \\ \hline
\multicolumn{5}{c}{Control} \Ts\\ \hline
yes \Ts& 3,566 &   841 & 28,459 & 1.24\% \\ 
no     & 5,594 & 1,011 & 20,012 & 0.88\% \\ \hline 
\multicolumn{5}{c}{Rumor} \Ts\\ \hline 
yes \Ts&  5,306 & 1,559 & 25,347 & 1.17\% \\ 
no     & 10,163 & 1,761 & 35,850 & 0.80\% \\ \hline
\end{tabular}
\end{table}

\item Finally, we compute the average tweeting rate for every user as the ratio of total number of lifetime tweets over the number of days since the account was created. To exclude what are likely to be automated accounts in both \n{rumor} and control datasets, we retain users with an average tweeting rate less than or equal to 24 tweets per day \n{(following posting activity thresholds such as in \cite{olteanu2017distilling,han2016cross})}. Applying this criteria, we discard 6,463 (26\%) users from rumor and 144,904 (31\%) users from control sets. 

\end{itemize}

For the remaining user accounts in both sets, we use the Twitter API user endpoint to collect the most recent 3,200 tweets, \n{synchronizing the time spans for the two datasets to span Paul \& Dredze timeline in 2012-2013}.


\n{\subsection{Relevance Refinement}}

\subsubsection{Human Labeling}

As thus far the data has been gathered using keyword matching, we refine the document (and thus, user) inclusion criteria by employing crowdsourced labeling and machine learning. In particular, we take this opportunity to make sure our data is on topic using CrowdFlower\footnote{\url{http://crowdflower.com/}} crowdsourcing platform to label a subset of the data, which then we use to build topical classifiers to determine the labels for the rest. \n{Note that instead of assessing the the veracity of the claims, we are now interested in making sure the text of the tweets indeed contains the cure claims, requiring basic lay language understanding, as is reflected in the task description below.}   

We begin by sampling the datasets. To ensure representativeness, we stratify the sample of rumor dataset such that at least 10 tweets from each topic are present, and the rest of the larger topics are sampled until a maximum of 100 tweets. This results in 4,152 tweets (which were de-duplicated by cleaned text). Similarly, we sample 4,000 tweets from control set for labeling. 

To ensure high quality of annotations, for each subset, 30 tweets were labeled and used as a ``gold-standard''. Using these tweets with known labels, the annotators are first given a quiz, and thereafter tested in each task (wherein gold standards are hidden among other tweets). The annotator must pass the quiz and maintain at least 70\% accuracy throughout the labeling process for their work to be accepted. \n{A minimum of t}hree independent labels were collected for each tweet \n{to achieve a majority decision}, and trial tasks of 100 tweets each were first run. \n{A total of 184 annotators were selected by CrowdFlower, contributing a minimum of 60 annotations each. Due to this large number of participants we report \%-age agreement instead of Fleiss' kappa.}

The tasks themselves differed slightly between the data sources. For control we ask the workers to label each tweet on (i) whether it is about cancer, and if so, (ii) whether there is a personal (or friend/family) experience, (iii) whether there is a claim that something cures cancer, or (iv) whether some other cancer-related information is present. For rumor (recall these tweets also mention some treatment or remedy) we ask whether the tweet (i) is about some cancer remedy, and if so, whether there is (ii) a claim it helps with treating or curing cancer, (iii) prevents cancer, or (iv) debunks such a claim. Given the tasks had multiple selections, the agreement was relatively high at 78.7\% for control and 82.0\% for rumor. 
Note, as discussed earlier, \n{in the instructions to the labelers we emphasized looking for a claim that the remedy} \emph{treats or cures cancer}, not just a symptom, with several illustrative examples for clarity. 

The results of the labeling task for the control tweets were as follows: 2,890 were labeled as having information about cancer whereas 1,110 tweets were labeled as non-related to the cancer topic. From the 2,890 cancer related tweets, 1,632 (40\%) were about personal experience, 98 (2\%) were about cancer cure and 1,1160 (29\%) were about other cancer-related information (symptoms, awareness, prevention, causes, etc.). The results of the labeling task for rumor tweets were as follows: 2,564 tweets were about a cancer cure and 1,587 were not about a cancer cure. From the 2,564 tweets about the remedy, 1,791 (43\%) tweets claimed that the suggested treatment helped to cure cancer (claimed a rumor), 564 (13\%) tweets were about prevention and 209 (5\%) tweets were debunking the claim. 

\subsubsection{Classification}
Next, we train several logistic regression classifiers on the labeled tweets using 1,2,3-grams as features. \n{We train the classifiers on the labeled tweets, which we then apply to the rest to characterize each user's behavior. } 
Summaries of selection for the two datasets are below: 

\begin{itemize}

\item Rumor: (1) is the tweet about a cancer cure? yes: 12,685, no: 7,872. Out of cancer cure tweets, (2) what kind of information does it have? claiming a cure: 9,549, prevention: 2,850, debunking claims of cure: 285. We define Rumor users as \textit{users who claim a cure is helpful for curing or treating cancer and \textbf{not} users who talk about prevention or debunking}, resulting in 12,046 tweets for 7,221 users.

\item Control: (1) is the tweet about cancer? yes: 339,047, no: 50,670. Out of cancer tweets, (2) which include a personal experience? yes: 197,608, no: 141,439. Further, (3) is the tweet is suggesting a cure? 
(Applying \textit{Synthetic Minority Oversampling Technique} \cite{chawla2002smote} to balance classes) cure: 2,252, not 336,794. We define Control users as \textit{users who post at least once about cancer, but \textbf{not} about a cancer cure}, resulting in 341,157 tweets for 270,622 users without and 199,343 tweets for 163,261 users with personal experience with cancer.

\end{itemize}

The overall process of user selection is summarized in Figure \ref{figuredatarefinement}, with resulting 16M tweets for 7,221 users in Treatment and 506M tweets for 443,883 users in Control datasets. Note that we do not make the distinction in the Rumor set between personal and non-personal experiences, as in a separate crowdsourced evaluation we find only 4\% to be about personal experiences.



%% file: results.tex
\section{Results}
\label{results}


\begin{table}[t]
\centering
\caption{Top rumor topics by number of unique users contributing tweets matching the expanded query.}
\label{tbl:toprumors}
{\fontsize{8pt}{9pt}\selectfont
\begin{tabular}{rlrrp{7cm}}
\hline
\textbf{\#} & \textbf{Topic} & \textbf{Users} & \textbf{Tweets} & \textbf{Expanded Query} \\\hline
70 & Juicing\Tf & 6,656 & 13,083 & juice OR juicing OR ``juice diet'' OR ``juice plus'' OR ``juice +'' OR ``fruit vegetable juice'' \\
11 & Apitherapy & 3,330 & 7,905 & apitherapy OR honey OR pollen OR ``bee bread'' OR ``propolis'' OR ``royal jelly'' OR ``bee venom'' OR ``bee sting'' \\
52 & Ginger & 3,113 & 8,928 & ginger \\
10 & Antioxidants & 2,908 & 5,671 & antioxidant \\
121 & Urine therapy & 2,532 & 4,686 & urine OR urinotherapy OR uropathy OR ``auto-urine therapy'' OR shivambu \\
9 & Antineoplaston therapy & 2,365 & 7,889 & antineoplaston OR burzynski \\
81 & Magnetic therapy & 2,327 & 30,789 & magnetic OR magnet OR magnets OR magnotherapy \\
124 & Walnuts & 2,013 & 5,474 & walnut OR walnuts OR ``Juglans regia'' OR akhrot OR ``wall nut'' \\
4 & Acupuncture & 1,817 & 5,359 & acupuncture OR accupuncture \\
103 & Poly-MVA & 1,705 & 7,252 & ``lipoic acid mineral complex'' OR ``poly-mva'' OR ``poly mva'' OR ``minerals vitamins and amino acids'' OR vitalzym OR curcumin OR ahcc OR essiac\\
\hline
\end{tabular}}
\end{table}

\begin{figure*}
    \centering
    \includegraphics[width=0.22\linewidth]{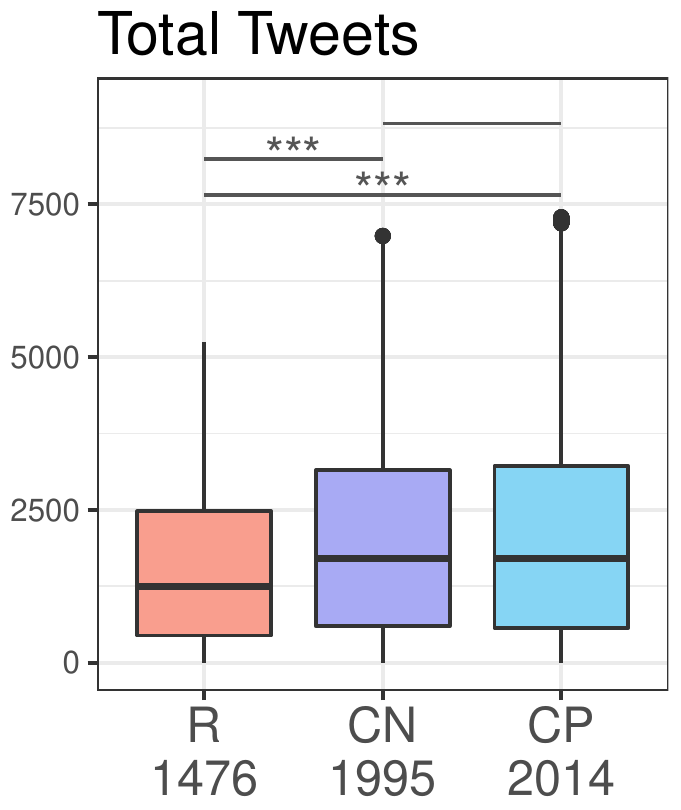}
    \includegraphics[width=0.22\linewidth]{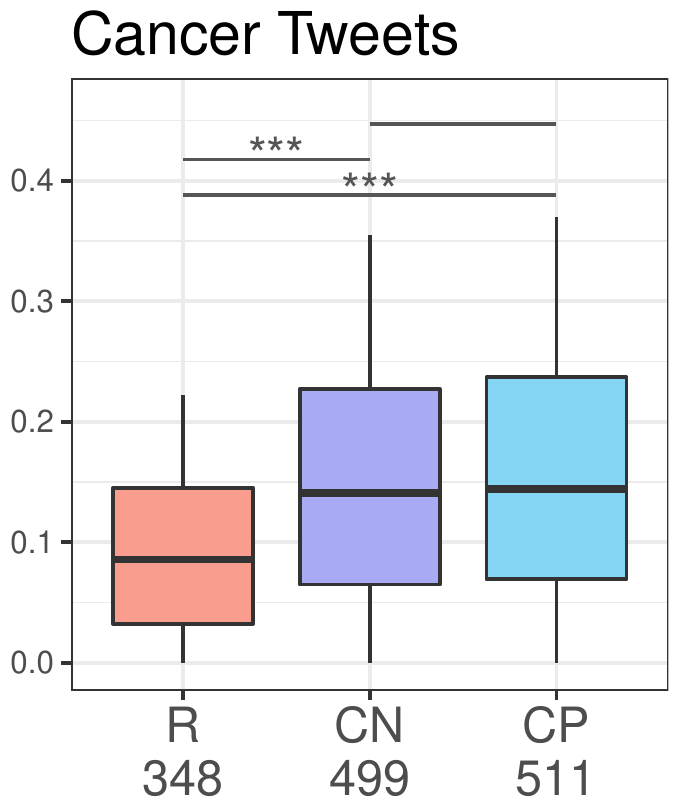}
    \includegraphics[width=0.22\linewidth]{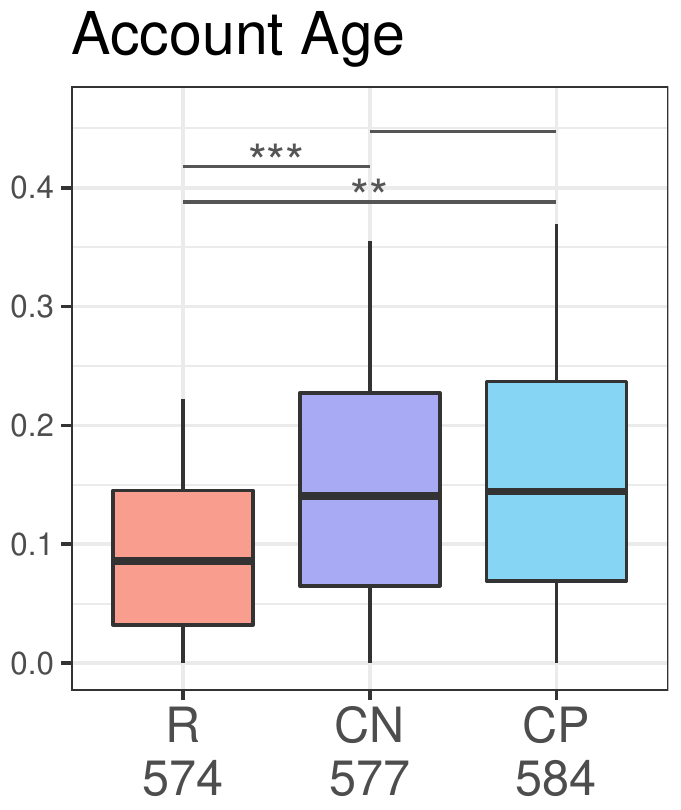}
    \includegraphics[width=0.22\linewidth]{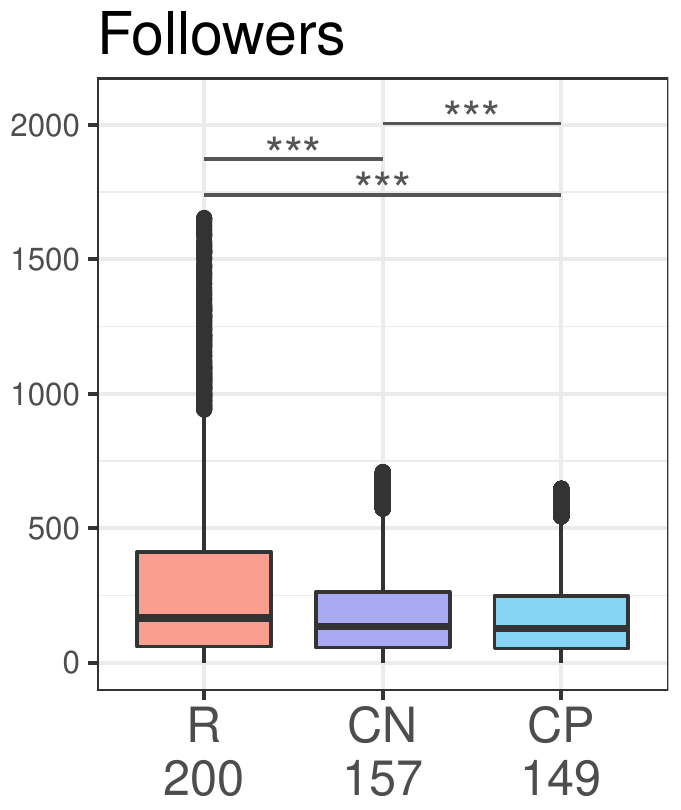}\\\vspace{0.3cm}
    \includegraphics[width=0.22\linewidth]{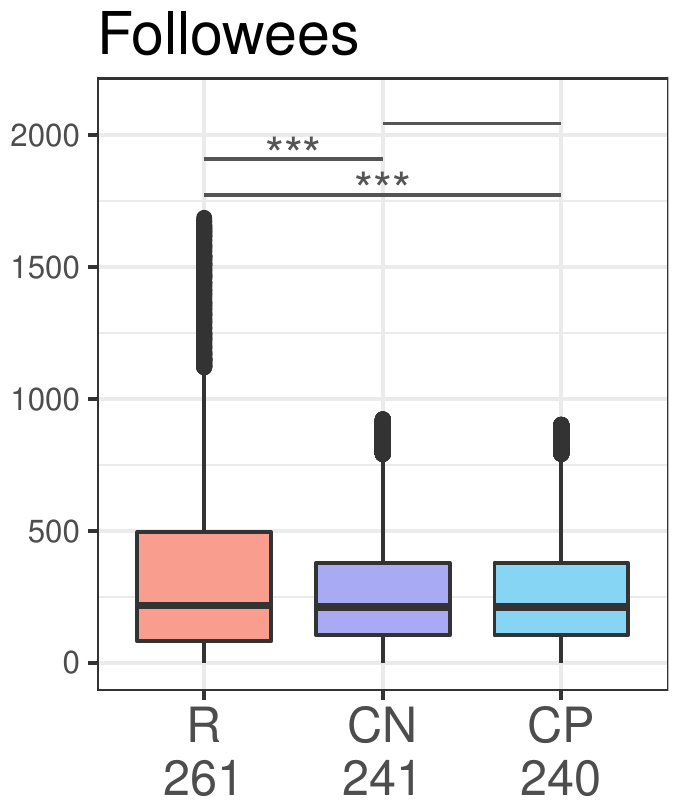}
    \includegraphics[width=0.22\linewidth]{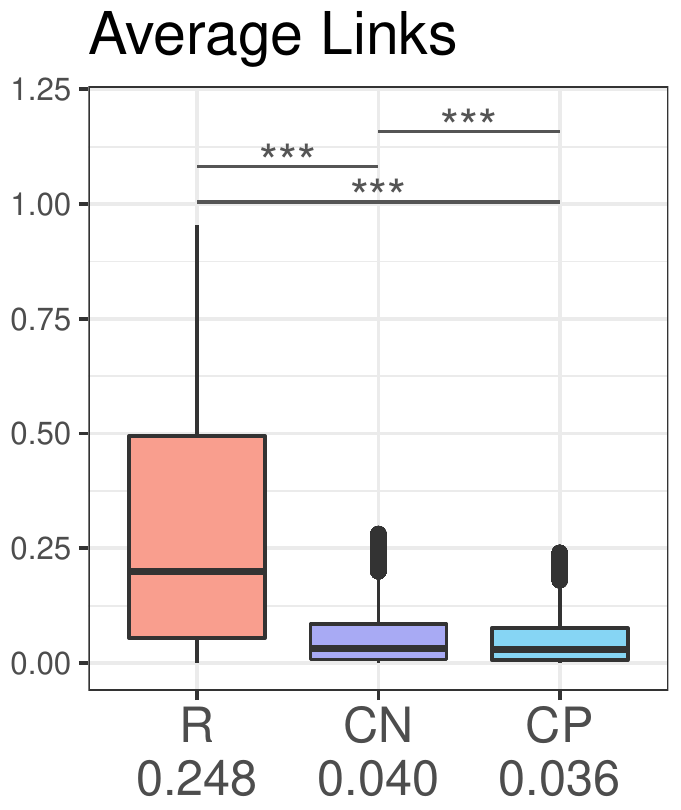}
    \includegraphics[width=0.22\linewidth]{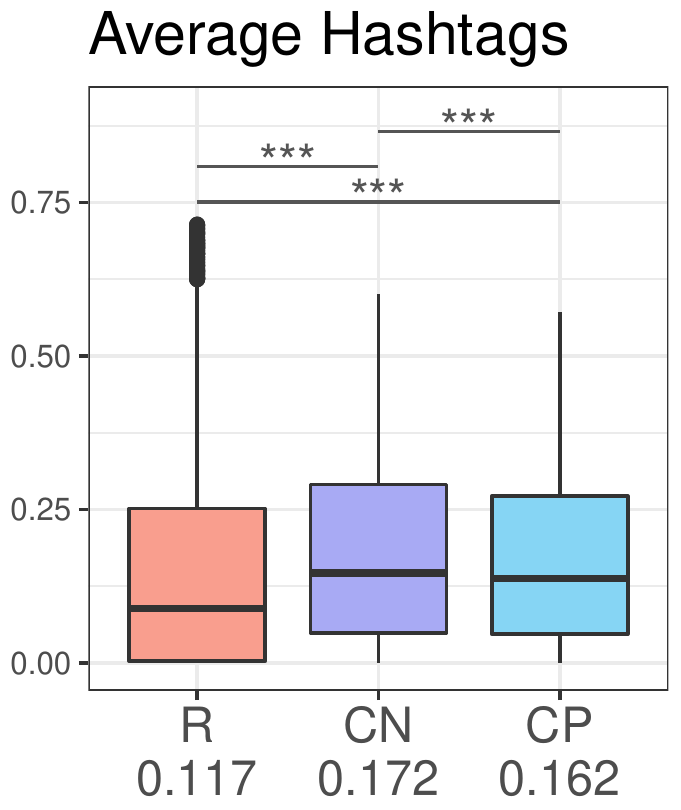}
    \includegraphics[width=0.22\linewidth]{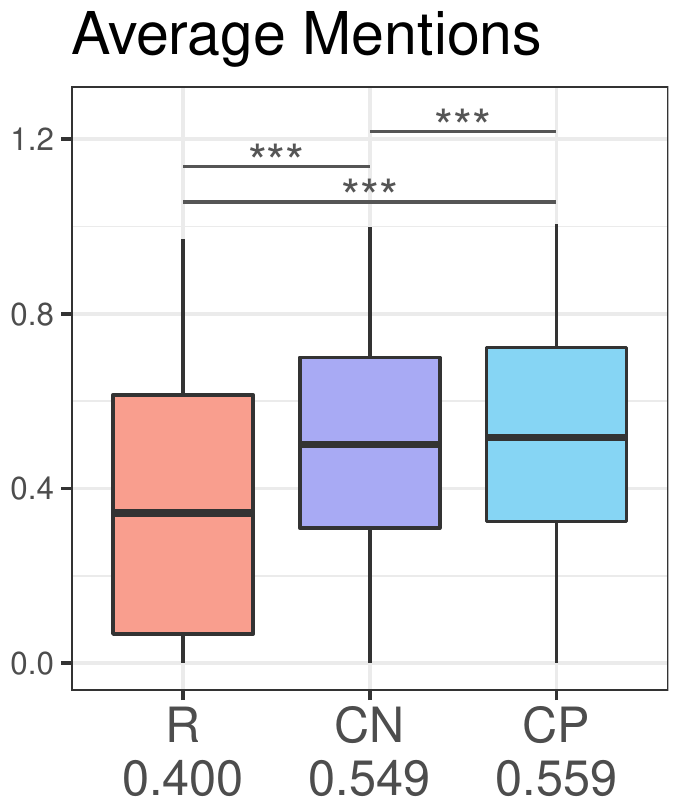}
    
    \caption{Summary of characteristics of Rumor, Control Non-personal and Control Personal user groups. For each characteristic a box plot (excluding outliers outside 90th percentile) is shown with median values under the title. Differences in medians are tested using Mann-Whitney U test, for which $p$-values\n{, Bonferroni adjusted for multiple hypothesis testing,} are shown on the corresponding lines spanning the two variables being compared: $p<0.0001$ ***, $p<0.001$ **, $p<0.01$ *.}
    \label{fig:boxplotstats}
\end{figure*}

\subsection{Rumor Topics}

Table \ref{tbl:toprumors} shows the ``treatments'' (or ``rumors'') which have the most user membership, along with the expanded queries which were used to collect the tweets. The most popular is \emph{juicing}, followed by similar widely available remedies, \emph{honey} and \emph{ginger}, as well as the \emph{antioxidant} keyword (which is often applied to a range of foods). We find a wide variety of claims surround foods and drinks. Some make bold claims outright: \emph{``[...] University show that the soursop fruit kills cancer cells effectively, particularly prostate cancer cells, pancreas and lung''}, others speculate \emph{``Can ginger help cure ovarian cancer?  Since 2007, the University of [...] has been studying GINGER... <url>''}, yet others invoke religious backing: \emph{``RT @<user>: Islamic backed \#cancer cure: Prophetic medicine cures woman of cancer using [...]: <url>''}. More unusual topics include \emph{Antineoplaston therapy} available in Dr. Burzynski clinic (for more, see Discussion), and \emph{urine therapy}. Note that the keyword queries returned both outrageous claims as well as debunking, such as \emph{``RT Dr. Burzynski ~ He has the cure for cancer, the FDA want to shut him down <url>''} and on the other side \emph{``Burzynski Clinic libel threat to silence critics of fake cancer treatment <url> @<user>''}\footnote{\n{Tweets have been slightly re-phrased to preserve user's privacy.}}. In Data section, we describe how we apply supervised machine learning to sort out actual claims of purported cures (from unrelated content or rumor debunking, for instance), and use them to identify the users engaged in misinformation. Overall, of the 139 topics collected, median number of tweets collected was 269.5, with a minimum of 11. These topics exemplify the breadth of the subjects covered in this dataset, as well as indicate the alternative cancer medicines popular on social media.

\subsection{Modeling Rumormongering}

We begin by comparing the users who have posted on these and other topics. Figures \ref{fig:boxplotstats} show box plots of behavioral statistics for the three kinds of users identified above (with outliers beyond 90th percentile excluded for clarity), such that the median is shown graphically as the bold line in each box, and also shown numerically under the label. The datasets are compared using using Mann-Whitney U test -- a non-parametric test that is more appropriate for highly skewed data for which normality cannot be assumed -- in the bars above the plots, with $p$-value level indicated symbolically. We find the Rumor user set to be quite different from the other two sets of users, having fewer total account lifetime tweets (1,476 compared to around 2,000 for Control), as well as cancer-related tweets in our dataset (more than 100 fewer), more followers and followees (some users being vastly more popular, note the long tails), and sharing more links (however fewer hashtags and mentions). Interestingly, in some behaviors there is a significant difference between personal and non-personal control tweets, with users having personal interactions with cancer having fewer followers, sharing fewer links and hashtags, but posting more mentions than non-personal control.

\begin{table}[t]
\centering
\caption{User and aggregated tweet features. }
\label{tbl:features}
{\fontsize{8pt}{9pt}\selectfont
\begin{tabular}{|l|l|l|}
\hline
\textbf{ Scope } \Tf      & \textbf{Feature}   & \textbf{Description}                                 \\ \hline \hline
{User} \Tf   & FOLLOWING          & The number of people the user is following                        \\ 
             & FOLLOWERS          & The number of people following the user                           \\ 
             & STATUS\_COUNT      & The number of tweets at posting time                              \\ 
             & ACCOUNT AGE        & The time passed since the author registered his/her account, in days \\
             & VERIFIED           & Whether account has been verified by Twitter                      \\
{Sentiment}\Tf & SENTIMENT SCORE  & Sentiment score value \cite{lowe2011scaling}                      \\
             & POSITIVE/NEGATIVE WORDS & The number of positive/negative words in text                \\     
             & EMOTICONS POS/NEG  & Count total number of positive and negative emoticons in text     \\
{Linguistic}\Tf & IS RETWEET      & Is a retweet; contains RT                                         \\
             & HAS MENTIONS       & Mentions a user, eg: @CNN                                         \\ 
             & HAS HASHTAG        & Contains hash\_tags                                               \\ 
             & URLS COUNT         & Count total number of URLs in text                                \\  
             & HASHTAG COUNT      & Count total number of hashtags                                    \\
             & MENTION COUNT      & Count total number of mentions                                    \\
             & WORD COUNT         & Count total number of words in text                               \\ 
             & CHAR COUNT         & Count total number of characters in text                          \\ 
             & UPPER COUNT        & Count total number of upper case letters                          \\ 
             & COUNT SENTENCES    & Count number of sentences                                         \\
             & QUESTION MARK      & Contains question mark '?'                                        \\
             & EXCLAMATION MARK   & Contains exclamation mark '!'                                     \\
             & PERCENTAGE UPPER/LOWER & The percentage of upper and lower case characters             \\ 
             & MULTIPLE QUES/EXCL & Contains multiple questions or exclamation marks                  \\ 
             & COUNT NOUN         & Count total number of nouns in text                               \\
			 & COUNT ADVERB       & Count total number of adverbs in text                             \\
             & COUNT ADJECTIVE    & Count total number of adjectives in text                          \\
             & COUNT VERB         & Count total number of verbs in text                               \\
             & COUNT PRONOUN      & Count total number of pronouns in text                            \\
             & HAS PRONOUN 1      & Contains a personal pronoun in 1th person                         \\ 
             & HAS PRONOUN 2      & Contains a personal pronoun in 2nd person                         \\ 
             & HAS PRONOUN 3      & Contains a personal pronoun in 3rd person                         \\     
             & LIWC               & 73 categories from psycholinguistic resource LIWC                 \\    
{Readability}\Tf & COMPLEX WORDS  & Count total number of complex words in text                        \\
             & READABILITY SCORES & Automated, Flesch\_Kincaid, Gunning, and SMOG \cite{readabilityscore}\\
             & COUNT NOT WORD2VEC & Count total number of words not in ``word2vec'' Google News vocabulary \\
             & AVG SYLLABLES      & The Average number of syllables per word in text                  \\
{Medical}\Tf & MEDICAL\_DOMAINS   & Refers to URL from a known medical organization, from (anonymized) \\ 
{Timing}\Tf  & INTERVAL ENTROPY   & Entropy of hour intervals between tweets from \cite{ghosh2011entropy}\\ 
                             \hline
\end{tabular}}
\end{table}

To examine the user behavior more deeply, we characterize the content which may be predictive of rumormongering behavior. In particular, we are interested in examining the tweets \emph{before} a user started posting about a rumor, not necessarily the claims themselves. Thus, for Rumor users we select the tweets before the first rumor post, and for Control we sample such a date from a normal distribution having mean and variance of first rumor posts of Rumor data. This way we aim to avoid biasing the selection to different time periods which may trivially differentiate users. This selection allowed for the analysis of at least 100 posts for 4,212 Rumor users.

\begin{table}[t]
\centering
\caption{Logistic regression with LASSO regularization model, predicting whether a user posts about a rumor, with forward feature selection. For each feature, coefficient (unstandardized), standard error, and accompanying p-value are shown. Significance levels: $p<0.0001$ ***, $p<0.001$ **, $p<0.01$ *, $p<0.05$ .}
\label{tbl:regression}
{
\selectfont
\begin{tabular}{lrrl}
\hline
\textbf{variable} & \textbf{coefficient} & \textbf{std. error} & \textbf{p-value} \\\hline
(Intercept)\Tf & -6.160 & 1.405 & *** \\
Avg syllables per word & 17.120 & 0.660 & *** \\
Is verified & -40.310 & 42310 &  \\
Percentage uppercase / lowercase & -0.201 & 0.018 & *** \\
Word count & 1.491 & 0.131 & *** \\
SMOG readability score & -0.753 & 0.123 & *** \\
Percentage uppercase & 0.191 & 0.019 & *** \\
Character count & -0.163 & 0.024 & *** \\
Number of cancer tweets & 0.001 & 1.9E-04 & *** \\
LIWC48: ingest & 1.839 & 0.722 & * \\
Negative word count & -1.460 & 0.262 & *** \\
URL count & 3.364 & 0.505 & *** \\
Is retweet & 4.947 & 0.790 & *** \\
word2vec count & -0.634 & 0.165 & *** \\
LIWC55: focuspast & -1.636 & 0.567 & ** \\
LIWC37: tentat & 2.531 & 0.859 & ** \\
Number of sentences & -0.610 & 0.205 & ** \\
LIWC32: male & -1.820 & 1.000 & \\
Interval entropy & 0.508 & 0.105 & *** \\
Account age & -0.001 & 2.7E-04 & *** \\
LIWC23: posemo & -0.490 & 0.384 &  \\
LIWC61: time & -1.431 & 0.378 & *** \\
LIWC13: adverb & 1.758 & 0.536 & ** \\
LIWC20: number & 2.936 & 1.317 & * \\
Statuses count & 7.1E-05 & 2.6E-05 & ** \\
LIWC42: hear & -4.742 & 1.799 & ** \\
Has 1st person pronoun & -1.504 & 0.662 & * \\
LIWC62: work & 1.591 & 0.665 & * \\
LIWC40: percept & 1.217 & 0.754 & \\ \hline
\end{tabular}}
\end{table}

Building on our previous work (anonymized), and use multifaceted behavior and content features which in the literature have been linked to credibility assessment of social media content. The features, listed in Table \ref{tbl:features}, span user-specific statistics, as well as aggregated (via averaging) tweet-specific metrics. User features encompass proxies of popularity (number of followers and followees), as well as productivity (number of posts up to date). Tweet features can be grouped into surface and linguistic forms of the tweet and well as semantically enriched ones including sentiment extracted from words and special characters, readability indices, and number of domains known to come from medical organization (anonymized). We also include a measure of entropy of the intervals between posts, which has been used to measure the predictability of retweeting patterns \cite{ghosh2011entropy}. Finally, we include the psycholinguistic resource LIWC\footnote{\url{https://liwc.wpengine.com/}}, which has been shown to relate to user mindset \cite{de2013predicting}.

We then turn to examining the relationship between these variables and the tendency of the user to post about a cancer treatment rumor. To mitigate class imbalance, we under-sample Control group by randomly sampling users to achieve a one to one balance with Rumor set. We then apply logistic regression with LASSO regularization, as the predicted class is binary and LASSO performs variable regularization and selection. However, as data has a large number of potentially collinear features, we also use forward feature selection method which employs Akaike Information Criterion (AIC) to select features contributing the most to the performance of the model \cite{venables2002modern} (note the significant features remain largely the same, but the selection process assists in ranking most prominent ones). The resulting model is shown in Table \ref{tbl:regression}, such that the features selected first are at the top. The McFadden R$^2$, the alternative to the R$^2$ of linear regression, is 0.925, indicating a good fit to the data. \n{We also perform a matched experiment wherein we match Rumor to Control users on the number of followers, such that for each Rumor user the closest match in Control is picked, resulting in McFadden R$^2$ of 0.906.} Examining the features, we can observe:
 
\begin{itemize}
\item We find \textbf{readability} to be of importance, with the average number of syllables per word and SMOG readability score at the top, as well as other style-related features. 
\item The fact whether or not account is \textbf{verified} is also important, however due to sparsity it is not statistically significant, indicating that such policing by the social media website may be of limited value. 
\item The top LIWC category is ``ingest'', one dealing with \textbf{eating and drinking}, echoing user's interest in topics potentially related to some of the most popular remedies we found (juices, superfoods, supplements, etc.).
\item These users are also more \textbf{prolific in writing about cancer}, with the number of cancer tweets being positively associated with posting a rumor (however each individual tweet counts little toward overall probability, with coefficient at 0.001). 
\item They are also more likely to use \textbf{tentative language} (LIWC category 37), possibly speculating about topics other than the rumors captured in this data.
\item Besides other LIWC categories pointing to speaking less positively and male and using more adverbs and numbers (as well as sharing more URLs), we find a weak negative relationship between using \textbf{first person pronouns} (``I'',``we''), indicating those engaging in posting about these rumors are not likely to be personally involved (remember also that we did not find many personal statements in Rumor set during labeling as well).
\item The positive relationship of posting \textbf{interval entropy} \cite{ghosh2011entropy} means the higher inter-tweeting time entropy -- and the less regular (not bot-like) is the posting behavior -- more likely the user to post about a rumor, pointing to a largely ``human'' cohort.
\end{itemize}

Thus, we find (likely non-bot) users who have a more sophisticated language, who are interested in cancer, and whose language already contains speculations (besides the rumor), but who are not personally involved with the illness.

To examine the language of these groups of users in more detail, in Table \ref{tbl:wordfreqs} we summarize the top 20 words, with stopwords removed, in all historical tweets by control users (left), all historical tweets of rumor users (center), and only rumor tweets (right). The frequency list on the right shows some of the main trends in the tweets explicitly endorsing a ``treatment''. Again, we find juices and antioxidants to be popular, and prominent mentions of ``help'', ``cure'', and ``treatment'' (with ``cure'' being the more popular keyword than ``treatment''). The center and the left lists show words in non-rumor tweets of Rumor users (center) and Control (left). Note that although both groups of users are in our dataset because at some point they have mentioned cancer, Rumor users are more focused on health, even when they are not explicitly talking about rumors, with these top 20 words containing 5 health-related words for Rumor users, and none in Control. Thus, we find an encouraging sign that propensity for posting cancer treatment misinformation can be modeled and predicted automatically. Next, we discuss ramifications of this observation.

\begin{table}[t]
\centering
\caption{\n{Word frequency tables summarizing the top 20 most popular terms, excluding stopwords, in all historical tweets by control users (left), all historical tweets of rumor users (center), and only rumor tweets (right).}}
\label{tbl:wordfreqs}
{\fontsize{8pt}{9pt}
\selectfont
\begin{tabular}{lrlr|lrlr|lrlr}
\hline
\multicolumn{4}{c}{Control History} & \multicolumn{4}{c}{Rumor History} & \multicolumn{4}{c}{Rumor Misinformation}  \\\hline
love & 1.95\% & night & 0.66\% & good & 1.01\% & video & 0.54\% & cancer & 1.43\% & cells & 0.50\% \\
good & 1.55\% & life & 0.63\% & health & 1.00\% & food & 0.54\% & juice & 0.81\% & out & 0.48\% \\
day & 1.34\% & happy & 0.60\% & day & 0.96\% & back & 0.50\% & RT & 0.77\% & healthy & 0.45\% \\
time & 1.22\% & ill & 0.59\% & love & 0.85\% & free & 0.46\% & breast & 0.73\% & diabetes & 0.44\% \\
people & 1.00\% & hope & 0.58\% & time & 0.78\% & work & 0.45\% & risk & 0.61\% & prostate & 0.44\% \\
lol & 0.99\% & feel & 0.55\% & great & 0.73\% & diet & 0.44\% & help & 0.58\% & antioxidant & 0.42\% \\
today & 0.96\% & haha & 0.51\% & people & 0.71\% & healthy & 0.40\% & health & 0.55\% & pain & 0.40\% \\
back & 0.94\% & follow & 0.51\% & today & 0.68\% & post & 0.38\% & helps & 0.54\% & chronic & 0.37\% \\
great & 0.73\% & home & 0.49\% & news & 0.62\% & weight & 0.38\% & cure & 0.54\% & patients & 0.37\% \\
work & 0.70\% & man & 0.47\% & life & 0.57\% & blog & 0.36\% & treatment & 0.53\% & study & 0.36\% \\
\hline
\end{tabular}}
\end{table}

%


%% file: discussion.tex
\section{Discussion}
\label{discussion}

\n{This study expands the misinformation research prominent in Social Computing, which has been largely focused on the political domain \cite{shu2017fake,chen2015misleading,gupta2013faking,ma2015detect,shao2016hoaxy}, to healthcare -- where erroneous beliefs and actions may cause serious bodily damage. Complementing HCI literature on human computer use and its sociocultural implications, this study extends current work on monitoring social media during crises and pandemics \cite{gui2017managing} as well as on the tracking of specific behaviors within a community of interest (such as in \cite{mejova2016fetishizing,almeida2016hci,mejova2017halal}). Below, we elaborate on the ecosystem of health communication and monitoring, possible application of our model, its theoretical contributions, and limitations.}


\n{\textbf{Context and Case Studies.}} Internet has long contributed to the ongoing ``deprofessionalization'' of medical practice. As Michael S. Goldstein writes in \emph{Persistence and Resurgence of Medical Pluralism}, ``Health information on the Internet enhances the autonomy of those who are ill, demystifies the knowledge and practices of doctors, and increases overall skepticism about medicine'' \cite{goldstein2004persistence}. In the larger history of antipathy toward professionals and their monopoly on knowledge, social media presents a new venue for patients, consumers, and concerned citizens to network and share their experiences and knowledge. It brings many of the remedies traditionally associated with home and family to a social domain. In such networked setting, knowledge may spread and evolve. For instance, \cite{lau2011online} found that people tended to change their belief about a health topic when it did not concur with a majority of others. These findings emphasize the potential power of social and word-of-mouth (WOM) marketing, which could benefit both from positive messages and from controversy around the product \cite{kozinets2010networked}.



Such marketing of lay health information may be especially effective in vulnerable populations, those having difficulty accessing medical care, or having poor medical literacy. For instance, a drug called ``Laetrile'', also known as ``Amygdalin'' or ``Vitamin B17'', is popularly promoted in India as an anti-cancer remedy \cite{bhatnagar2017laetrile}. Despite a ban on its marketing as such by Food and Drug Administration (FDA) in the United States in 1979, it remains popular in India \cite{helen2008cancer}, and in our data we found 2,417 mentions of it in the context of cancer. Currently, Laetrile is promoted on YouTube\footnote{A YouTube search for ``laetrile'' on April 15, 2018 has resulted in a top video titled ``Learn How Laetrile Kills Cancer Cells!''.} and other social media -- media to which general public has much more access than scientific literature -- allowing for an international audience to be reached. An exciting future research direction lies in enriching our dataset with geo-location in order to track the supporters of these treatments across the world.

However, an opposite reaction can also be possible. A clinic purporting to cure its patients of cancer by its founder Stanislaw R. Burzynski, MD has for several decades been a subject of scientific renunciation \cite{green1992antineoplastons}. Recently, bloggers and activists which have been criticizing Burzynski for ``disturbing business and research practices'' have been attacked personally on social media, which encouraged these ``skeptics'' to organize and educate the public about the unproven nature of Burzynski's treatments \cite{blaskiewicz2016skeptic}. In 2017, Dr. Burzynski was placed on probation for five years by the Texas Medical Board and ordered to pay a total of \$60,000 in fines and restitution for not adequately informing patients about the treatments that they were receiving \cite{chang2017texas}. We find social media and Twitter specifically to be a new battleground for the health claims of different parties, some of which may be businesses set to lose profit if their message is contested. In our dataset, we found 7,889 tweets mentioning Burzynski or ``Antineoplaston therapy'' he proposes. Case studies of such two-sided interactions provide a window into the consequences of increased plurality in voices aiming to spread health-related information.

\n{\textbf{Applications.}} As beliefs are strongly linked to behavior, honing internet-enabled communication with patients and public at large is important in improving interventions of health behaviors. 
\n{In this context, present work contributes to the Social Computing community in proposing a tool for monitoring health misinformation on a large scale. Specifically, the model built in this study exemplifies specialized tools that can help address the spread of health misinformation on social media, mainly in (i) automatically detecting Twitter users who may be likely to post questionable information, (ii) attempting to change those individuals' view of the topic, and (iii) quickly identifying and limiting the spread of misinformation. }

\n{First, in the age of personalization, such models can be employed to target individuals potentially susceptible to follow questionable accounts, consume poor quality health information, and propagate it. Early identification of rumor-prone individual accounts allows for refining traditional broad-spectrum information campaigns via personalized employment of \emph{persuasive technologies} which offer a way to tailor content to the individual and track individual progress \cite{berkovsky2012influencing}. In particular, such technologies attempt to nudge the user to change his or her attitude or behavior through persuasion or social influence (but not misinformation or coersion), and have already been applied to health coaching and communication \cite{cugelman2013gamification,yardley2015person}.} 

\n{Secondly, multi-faceted features proposed in this study provide a foundation for examining both behavioral characteristics and interests of those prone to rumormongering. Such interest lists can be expanded beyond LIWC to specialized topical lexicons, such as those on vaccination hesitancy \cite{dunn2015associations}, eating disorders \cite{yom2012pro,ghaznavi2015bones}, antibiotics \cite{scanfeld2010dissemination}, etc.}  

\n{Finally, further monitoring of suspect accounts will allow for timely identification of new potentially questionable content before it has a chance to propagate through the network, alerting public health officials of new waves of content or public interest. This content then can be automatically pre-assessed for credibility using approaches such as in \cite{castillo2011information}, notifying officials if the content passes a certain threshold. 
}

\n{Note that, automated tools will not be able to replace expert knowledge, but instead contribute to a fruitful human-expert-in-the-loop paradigm which has been proposed for research and machine learning processes \cite{girardi2016interactive,holzinger2016interactive}. In particular, we describe a pipeline for training the model for the tracking of discussions around ``complementary and alternative medicines'' for cancer, and we show that it achieves a high McFadden R$^2$ in fitting the data, however the pipeline can be applied to any other healthcare topic. Further, in order to remain relevant in the changing discourse, it must be periodically re-trained with fresh data in order to ameliorate ``concept drift'' (for which streaming solutions are also being developed \cite{ghazikhani2014online}). 
}

\n{Thus, contributing to CSCW community's interest in application-driven computer supported systems, our tool is a potential component in the public health communication and monitoring ecosystem. For example, the largest regulator of possible cancer treatments in the US is the Food and Drug Administration (FDA), which solicits reports concerning ``defects in the quality or safety'' of some product via Safety Reporting Portal \cite{USFoodandDrugAdministration2018}. Such reporting scheme relies on parties existing who are concerned about a particular product, who to study it and file a report. Such system allows in the meanwhile for potentially dangerous health products and advise to affect the public. Outside official channels, social media is being explored by thinktanks in order to track adverse drug side effects \cite{belbey2016fda}, however to the best of our knowledge currently no official FDA social media surveillance systems exist. Early detection of harmful information is a necessary precondition to timely deployment of corrective communication, if not to the sources of misinformation, then to their immediate audience. As such, early detection systems contribute to the discussion of computer-supported emergency communication in CSCW community \cite{olteanu2015expect,leavitt2017role}. Thus as outlined above, automated detection of social media accounts susceptible to propagating health misinformation proposed here has great potential for expanding the capacity for preventative monitoring and communication of current healthcare communication ecosystem.}



\n{\textbf{Theory of Rumormongering.}} More generally, psychological underpinnings of believing and propagating rumors have been studied in the context of larger societal impact. In post WWII Europe, Gordon W. Allport and Leo Postman \cite{allport1947psychology} studied rumors out of concern about the damage to morale and national safety. In the ample literature following, rumors have been defined as public communications that are infused with private hypotheses about how the world works, in particular to help us cope with our anxieties and uncertainties \cite{rosnow1991inside}. Hypothesizing about the nature of rumor, Allport and Postman postulated that the strength of a rumor (R) will vary with the importance of the subject to the individual concerned (i) times the ambiguity of the evidence pertaining to the topic at hand (a), or $R \approx i \times a$~\cite{rosnow2005rumor}. \n{Contributing to this theoretical work, our empirical analysis of social media health misinformation shows that those engaged in spreading unproven cancer treatments largely are not personally involved in the matter. Thus, we propose to extend the definition of ``importance'' $i$ to other motivating factors beyond the personal, which may be at play in the public sphere of Twitter. Further analysis is required to reveal motivating factors in such health misinformation spread, as cancer fraud has been acknowledged in Oncology literature \cite{vogel2011internet}.}

Lessons learned from the empirical studies of psychological drives in rumormongering are essential in building more effective policies on communicating scientific information and managing public opinion on issues of medicine and related policies. \n{Unlike in political domain where bots can hijack the conversation \cite{shao2017spread,davis2016botornot}, we find that posting interval entropy (measuring irregularity of post timings) was positively related with a rumormongering behavior, pointing to a more ``human'' trait. This finding emphasizes the importance of public education and communication campaigns as preventive measures targeting public beyond social media.} For instance, after a public outcry in Italy to legalize a stem cell-based treatment for neurological diseases unsupported by published evidence, researchers and public health officials called for an improvement in guidelines for its media on communicating scientific information to the public \cite{kamenova2015stem}. Special care needs to be taken to promote clarifications and retractions, as it has been shown that these are not as popular as, for example, the original wrong news stories \cite{eysenbach2005cybermedicine}.







\n{\textbf{Limitations}}. Studying health misinformation on social media has several important limitations. Social media adoption and use differs widely between population segments: for instance, close to half (45\%) of 18- to 24-year-olds in U.S. use Twitter, compared to 24\% of all adults, as reported by Pew Research Center in March 2018 \cite{PewResearchCenter2018social}. \n{Detecting legitimate personal accounts (as opposed to bots or organizational accounts) remains a challenge, which we attempted to address using existing tools like Humanizr and baby name dictionaries, which undoubtedly introduce their own biases potentially excluding certain minorities. In particular, the Social Security name database includes all names registered at least 5 times in a year, dating back to 1880, capturing a large majority of names used. However, more resources could have been used to include the names of minorities, such as the Register of Liberated Africans\footnote{\url{http://liberatedafricans.org/}}. The results of this study should be taken in the light of this limitation, as we may have failed to detect misinformation in some communities.  Improvement in the detection of real humans (versus bot or organizational accounts) will allow for a more accurate account selection for studying individuals.}

Accessibility issues also bias the view of the populations having, for instance, visual impairment \cite{wu2014visually} or other constraints to using the medium. Further, some health conditions and personal topics are associated with a social stigma which limit their discussion on social forums. For example, \cite{de2014seeking} found some illnesses to be searched more often than discussed in social media -- a bias which may affect our selection of certain cancers. Finally, attitudes toward self-expression and trust in publicly available information may differ wildly between cultural subgroups, such as in case of Hong Kong youths, who were found to be significantly more likely to disclose personal health issues with peers online compared to their U.S. counterparts \cite{lin2016health}. Hong Kong youths also held the highest level of trust towards health-related information on social media, again pointing to the need of personalized approach to health communication sensitive to the culture of the participants. Finally, observations in this study concern exclusively treatment claims of cancer, and may not generalize to other illnesses, especially if they have different societal stigmas. \n{The model proposed in this work inherits the above limitations, thus any integration of such automated tracking must be closely monitored for bias and topic drift, and regularly updated in order to capture latest developments in social media norms.}

Finally, as any technology, the proposed analytical pipeline may be misused when applied within faulty policies. The expert definition of misinformation must be subjected to ethical constraints of medical and public health standards. How the information that is flagged by the system is handled must also avoid discouraging public discourse and information seeking.

\textbf{Privacy}. This study used Twitter posts which were publicly available at the time of data collection, with no private messages or messages deleted by the time of the collection included. Also, accounts which have been deleted since Paul \& Dredze collection have not been included in the study. Furthermore, the sharing of this dataset will be done according to Twitter's Terms of Use.

%% file: conclusion.tex
\section{Conclusion}
\label{conclusion}

In this paper we present a case study of health misinformation on social media by examining Twitter users involved in propagating alternative medicines claiming to treat or cure cancer. Through a multi-stage process including machine learning, crowdsourcing, and heuristics, we select users who are likely to be real people, and who post on one of 139 such topics. We find that these users are likely to use more sophisticated language, and circulate in health domain prior to posting a rumor, but are not likely to be personally involved in the illness. Our findings suggest that cancer treatment misinformation may be spread not by patients, but by other actors. More research needs to be done to ascertain motivations, tactics, and impact of such accounts. Finally, the dataset collected for this study presents a highly-curated resource for the research community's future studies on the topic of health misinformation.